\documentclass[twocolumn,showpacs,preprintnumbers,amsmath,amssymb,prb]{revtex4}
\usepackage[dvips]{color,graphics,epsfig,rotating}
\usepackage{graphicx}% Include figure files
\usepackage{dcolumn}% Align table columns on decimal point
\usepackage{bm}% bold math

\begin{document}

\title{Renormalized behavior and proximity to a magnetic quantum
critical point in BaCo$_2$As$_2$}

\author{A.S. Sefat, D.J. Singh, R. Jin, M.A. McGuire, B.C. Sales, D. Mandrus}
\affiliation{Materials Science and Technology Division,
Oak Ridge National Laboratory, Oak Ridge, Tennessee 37831-6114}

\date{\today}

\begin{abstract}
We report synthesis and single crystal measurements of magnetic,
transport and thermal properties of single crystalline
BaCo$_2$As$_2$ as well as first principles calculations of the
electronic structure and magnetic behavior. These results show
that BaCo$_2$As$_2$ is a highly renormalized paramagnet in
proximity to a quantum critical point, presumably of ferromagnetic
character and that BaFeNiAs$_2$ behaves similarly. These results
are discussed in relation to the properties of Ba(Fe,Co)$_2$As$_2$
and Ba(Fe,Ni)$_2$As$_2$, which are superconducting for low Co and
Ni concentrations.
\end{abstract}


\maketitle

The discovery of high temperature superconductivity in
oxy-pnictide phases, prototype LaFeAs(O,F) by Kamihara and
co-workers \cite{kamihara} has led to strong interest in
establishing the physical properties of these materials and the
mechanism of superconductivity. Since the discovery,
superconductivity has been found in a rather wide variety of
compounds with Fe$^{2+}$ square planar sheets. These include the
oxy-pnictides, doped ThCr$_2$Si$_2$ structure compounds of e.g.
BaFe$_2$As$_2$, \cite{rotter2} LiFeAs, \cite{wang} and
FeSe.\cite{hsu} Remarkably, superconductivity can be produced in
BaFe$_2$As$_2$ by alloying Fe with the other ferromagnetic 3$d$
elements, Co or Ni, \cite{sefat,li} as is also found in the
oxy-pnictides. \cite{sef-1} Interestingly, both the Ni-based
oxy-pnictide and BaNi$_2$As$_2$ are superconductors,
\cite{ronning,watanabe,li-ni} although the critical temperatures
are much lower than in the Fe-based compounds and the mechanism
may be different. \cite{sub1,sub2} Also, LaCoAsO and LaCoPO are
itinerant ferromagnets. \cite{yanagi,sef-1}

The single crystals of BaCo$_2$As$_2$ were grown out of CoAs flux.
The typical crystal sizes were $\sim$ 5x3x0.2 mm. High purity
elements ($>$ 99.9 \%, from Alfa Aesar) were used in the
preparation of the crystals. First, CoAs binary was prepared by
placing mixtures of arsencic pieces, and cobalt powder in a silica tube. These were
reacted slowly by heating to 300$^\circ$C (dwell 10 hrs), to
600$^\circ$C (dwell 10 hrs), then to 900$^\circ$C (dwell 10 hrs).
Then, a ratio of Ba:CoAs = 1:5 was heated in an alumina crucible
for 15 hours at 1180$^\circ$C under a partial atmosphere of argon.
This reaction was cooled at the rate of 2$^\circ$C/hour, followed
by decanting of CoAs flux at 1090$^\circ$C. The crystals were
well-formed plates with the [001] direction perpendicular to the
plane. Electron probe microanalysis of a cleaved surface of the
single crystal was performed on a JEOL JSM-840 scanning electron
microscope using an accelerating voltage of 15 kV and a current of
20 nA with an EDAX brand energy-dispersive X-ray spectroscopy
(EDS) device. EDS analyses indicated Ba:Co:As ratio of 1:2:2. The structural identification was made via powder
x-ray diffraction, using a Scintag XDS 2000 $\Theta - \Theta$
diffractometer (Cu K$_\alpha$ radiation). The cell parameters were
refined using least squares fitting of the measured peak positions
in the range 2$\Theta$ from 10 - 90$^\circ$ using the Jade 6.1 MDI
package. The lattice parameters of BaCo$_2$As$_2$ are $a$ =
3.954(1) \AA, and $c$ = 12.659(3) \AA, in the ThCr$_2$Si$_2$
structure ($I4/mmm$, $Z$ = 2) in close agreement with the report
of Ref. \onlinecite{pfisterer}.

\begin{figure}[tbp]
 \includegraphics[width=0.82\columnwidth]{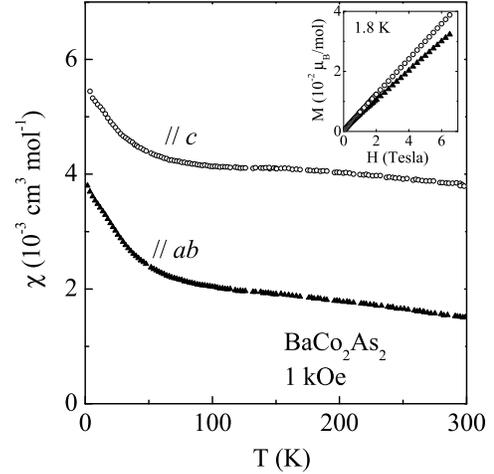}
 \caption{Magnetic measurements for BaCo$_2$As$_2$
along the two crystallographic
directions. The figure is the temperature dependence of the zero-field
cooled magnetization in 1 kOe; inset is the field dependence of the magnetization at 1.8 K.}
\label{Fig1}
\end{figure}

DC magnetization was measured as a function of temperature and
field using a Quantum Design Magnetic Property Measurement System.
Fig. \ref{Fig1} shows the susceptibility in 1 kOe applied field
along $c$- and $ab$-crystallographic directions. The
susceptibility is weakly anisotropic and decreases with increasing
temperature. At 1.8 K, $\chi_{c}$ = 5.4 x 10$^{-3}$
cm$^{3}$ mol$^{-1}$ and $\chi_{ab}$ = 3.8 x 10$^{-3}$ cm$^{3}$ mol$^{-1}$.
As seen in the inset of Fig. \ref{Fig1} the field dependent
magnetization is linear at 1.8 K.

\begin{figure}[tbp]
 \includegraphics[width=0.95\columnwidth]{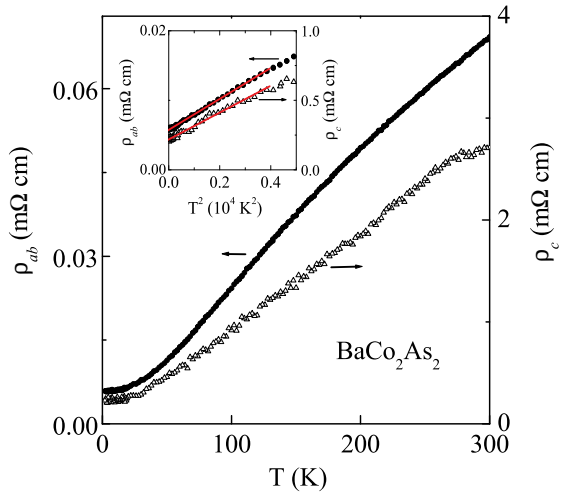}
 \caption{Temperature dependence of the resistivity of BaCo$_2$As$_2$
in zero field for the two crystallographic directions.}
\label{NewFig2}
\end{figure}

Temperature dependent electrical resistivity measurements were
performed on a Quantum Design Physical Property Measurement System
(PPMS), measured in both the $ab$ plane ($\rho_{ab}$) and $c$ direction ($\rho_c$).  As shown in Fig. \ref{NewFig2}, both $\rho_{ab}$ and $\rho_c$ decrease with decreasing temperature, revealing metallic behavior. However, the
resistivity is much more anisotropic than the magnetic susceptibility.  At room temperature
$\rho_{ab}$ = 0.07 m$\Omega$ cm and $\rho_c$ = 2.7 m$\Omega$ cm, resulting in $\rho_c$/$\rho_{ab}$ $\sim$ 39.  In analyzing the low temperature data, we found that both $\rho_{ab}$ and $\rho_c$ exhibit quadratic temperature dependences in a wide temperature range.  Shown in the inset of Fig. \ref{NewFig2} is the plot of $\rho_{ab}(T^2)$ and $\rho_c(T^2)$ between 1.8 and 71 K.  Note that $\rho_{ab}$ and $\rho_c$ vary approximately linearly with T$^2$ below $\sim$ 60 K.  By fitting the resistivity data between 1.8 and 60 K using $\rho_{ab,c}$ = $\rho_{ab,c}$(0 K) + \textit{A}T$^2$, we obtain the residual resistivity $\rho_{ab}$(0 K) = 5.75 $\mu\Omega$ cm, $\rho_c$(0 K) = 0.22 m$\Omega$ cm, and constant \textit{A}$_{ab}$ = 2.22$\times$10$^{-3}$ $\mu\Omega$ cm/K$^2$, \textit{A}$_c$ = 9.65$\times$10$^{-2}$ $\mu\Omega$ cm/K$^2$.  These give
the residual-resistivity-ratio ($\rho_{\rm 300 K}/\rho_{\rm 0 K}$) $\sim$ 12
along both crystallographic directions, indicating
good crystal quality. The quadratic temperature dependence of the electrical resistivity reflects the importance of the Umklapp process of the electron-electron scattering at low temperatures and is consistent with the formation of a Fermi-liquid state.  Interestingly, the system shows little magnetoresistance under application is magnetic field up to 8 Tesla.

\begin{figure}[tbp]
 \includegraphics[width=0.84\columnwidth]{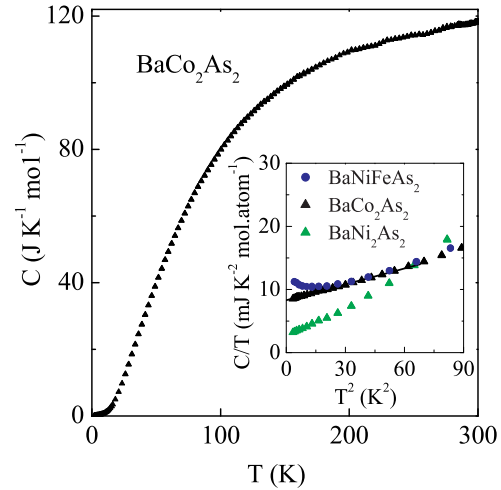}
 \caption{(Color online)
Temperature dependence of the specific heat of BaCo$_2$As$_2$.
The inset is $C/T$ vs. $T^2$ and a linear fit to the data below $\sim$7 K
compared with data for BaFeNiAs$_2$ and BaNi$_2$As$_2$.}
\label{NewFig3}
\end{figure}

Specific heat data, $C_p(T)$, were also obtained using the PPMS
via the relaxation method. Fig. \ref{NewFig3} gives the temperature
dependence of specific heat. Below $\sim$ 7 K, $C/T$ vs. $T^2$ is
linear (inset of Fig. \ref{NewFig3}) consistent with a Fermi liquid
plus phonon contribution. In the temperature range of 1.9 K to
$\sim$ 7 K, the fitted Sommerfeld coefficient, $\gamma$, is 8.2
mJ/(K$^2$ mol atom) or 20.5 mJ/(K$^2$ mol Co). This is a high
value that would correspond to an electronic density of states of
8.7 eV$^{-1}$ per Co both spins. The mean field Stoner criterion
($NI > 1$, $I \sim$ 0.8 eV per spin for 3\textit{d} elements) for
ferromagnetism would be strongly exceeded if this is interpreted
as the bare electronic density of states for Co. The absence of
magnetism therefore implies either strong hybridization with
ligands, which is not evident in the electronic structure, or
substantial renormalization due to e.g. spin fluctuations.
Assuming that $\chi_{c}$ = 5.4 x 10$^{-3}$
mol$^{-1}$ and $\chi_{ab}$ = 3.8 x 10$^{-3}$ mol$^{-1}$ at 1.8 K
are spin susceptibilities (i.e. assuming that the diamagnetic and
van Vleck contributions are small),
we may estimate the Wilson ratio
R$_W$=$\pi^2k_B^2\chi_{spin}/(3\mu_B^2\gamma$).
This gives R$_W$ $\sim$7 from $\chi_{ab}$ and $\sim$10 from $\chi_c$.
These values well exceed unity for a free electron system and are
indicative of enhanced magnetic behavior.
There are several possible reasons for high R$_W$:
one is that actual low-temperature spin susceptibility is much smaller
than the total susceptibility e.g. due to a strong orbital
contribution.
Another source of enhancement can be strong Coulomb correlations, which often
result in Wilson ratios intermediate between 1.5 and 2.0.
However, the large value that we find, which is an order of magnitude
higher, is best explained as indicating that the system is close to
ferromagnetism.
Combined with the high R$_W$, we find
a normal Kadowaki-Woods ratio \textit{A}/$\gamma^2$.
In particular, for BaCo$_2$As$_2$ in-plane transport we obtain
\textit{A}$_{ab}$/$\gamma^2$ = 0.5$\times$10$^{-5}$
$\mu\Omega$ cm/(mJ/mol-K)$^2$. This is comparable to but
larger than the value for the nearly
two dimensional Fermi liquid, Sr$_2$RuO$_4$,
somewhat smaller than the typical value
of $10^{-5}$ $\mu\Omega$ cm/(mJ/mol-k)$^2$ for heavy Fermions,
and significantly smaller than the values found in correlated oxides with
frustration or other very strong scattering such as LiV$_2$O$_4$ or
La$_{1.7}$Sr$_{0.3}$CuO$_4$. \cite{li}

\begin{figure}[tbp]
 \includegraphics[height=2.8in,angle=270]{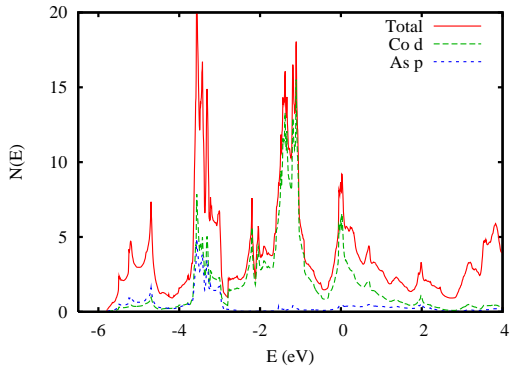}
 \caption{LDA non-spin-polarized electronic density of states of BaCo$_2$As$_2$
on a per formula unit basis. The projections shown are onto the LAPW spheres.}
\label{dos-ck}
\end{figure}

\begin{figure}[tbp]
 \includegraphics[width=2.8in]{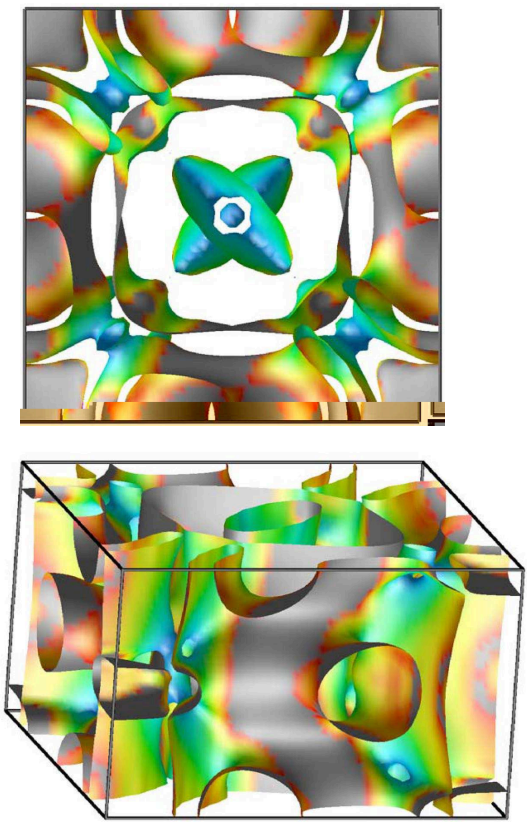}
 \caption{(Color online)
LDA non-spin-polarized Fermi surface of BaCo$_2$As$_2$
shaded by velocity; blue is low velocity. The
corners are $\Gamma$ and $M$ (0,0,1/2) points in the body centered
tetragonal zone. }
\label{fermi}
\end{figure}

Electronic structure calculations were performed using the general
potential linearized augmented planewave (LAPW) method with local
orbitals, \cite{singh-book,singh-lo} similar to those reported
previously for BaFe$_2$As$_2$. \cite{singh-ba} The calculations
used LAPW sphere radii of 2.2, 2.1 and 2.1 $a_0$ for Ba, Co and
As, respectively, and were done within the local density
approximation (LDA) with reported lattice parameters of $a$=3.958
\AA, and $c$=12.67 \AA. \cite{pfisterer} The As internal
coordinate $z_{\rm As}$ was obtained by energy minimization as
$z_{\rm As}$=0.34528. The calculated electronic density of states
(DOS) and Fermi surface are shown in Figs. \ref{dos-ck} and
\ref{fermi}.

The DOS is rather similar to that of the Fe compounds with the
exception of the position of the Fermi energy. In particular, it
shows only modest covalency between Co and As, and has a pseudogap
at an electron count of six $d$ electrons per Co. The extra $d$
electron in Co$^{2+}$ relative to Fe$^{2+}$ shifts $E_F$ upwards
into a region of high DOS. The Fermi surface is large and
multi-sheeted, as shown in Fig. \ref{fermi}. The calculated Fermi
surface has both two and three dimensional sheets, and in
particular is more three dimensional than might be expected based
on the anisotropy of the measured resistivity. Reconciling the
Fermi surface with the transport data would require scattering
that depends on the particular Fermi surface sheet, as may occur,
for example, from spin fluctuations with a large anisotropy or due
to strong band dependence of the electron phonon scattering. The
main As contributions to the DOS are between -6 eV and -3 eV
relative to $E_F$, corresponding to stable As$^{3-}$ and
Co$^{2+}$. The value at the Fermi energy is $N(E_F)$ = 8.5
eV$^{-1}$ on a per formula unit (two Co atoms) both spins basis.
Comparing with the measured specific heat implies a mass
renormalization of $\sim$ 2, similar to the Fe-based
superconductors. As shown in the inset of Fig. \ref{NewFig3}, our
measured value for BaNi$_2$As$_2$ is $\gamma$ = 15 mJ/(mol K$^2$),
as compared with a bare value from $N(E_F)$ of 8.9 mJ/(mol K$^2$).
This yields an enhancement factor of $(1+\lambda)=1.7$, consistent
with the value of the electron-phonon $\lambda$ = 0.76 obtained
from first principles calculations for that compound. \cite{sub2}

\begin{figure}[tbp]
 \includegraphics[height=0.90\columnwidth,angle=270]{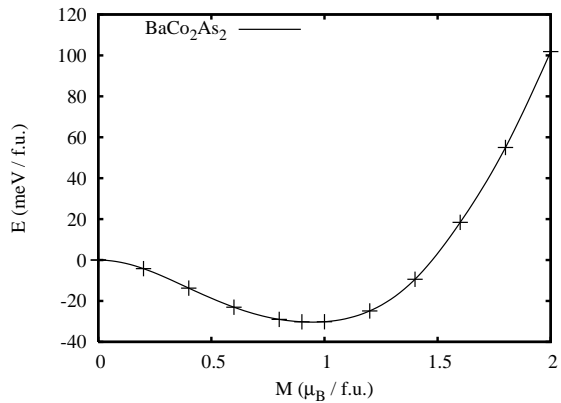}
 \caption{LDA fixed spin moment energy as a function of constrained
moment for BaCo$_2$As$_2$.}
\label{fsm}
\end{figure}

The physics, however, appears different in BaCo$_2$As$_2$. First
of all, the susceptibility implied by the magnetization
measurements is enhanced over the Pauli susceptibility from the
LDA density of states, $\chi_0$ = 2.75x10$^{-4}$ emu/mol by a
factor of $\sim$ 15 (depending on direction) yielding a high
Wilson ratio with implied nearness to ferromagnetism. Furthermore,
the non-renormalized LDA $N(E_F)$ is already large enough to lead
to a mean field Stoner instability towards ferromagnetism. In
fact, contrary to experiment we find a stable ferromagnetic ground
state in the LDA, with spin moment 0.94 $\mu_B$ per formula unit
(0.42 $\mu_B$/Co) and energy 32 meV per formula unit lower than
the non-spin-polarized solution. The total energy as a function of
constrained moment is shown in Fig. \ref{fsm}. We also searched
for an antiferromagnetic solution with a checkerboard in plane
arrangement where all nearest neighbor Fe bonds are
antiferromagnetic. However, we did not find a stable
antiferromagnetic state of this type. This indicates that the
stability of Fe moments depends on the particular magnetic order
and therefore that the magnetism is itinerant. Thus at the LDA
level, BaCo$_2$As$_2$ is a weak itinerant magnet, apparently
ferromagnetic in-plane, though the out-of-plane interaction may be
ferromagnetic or antiferromagnetic.

While the mechanism for
superconductivity of the layered Fe compounds
remains to be established, there is accumulating
evidence of a relationship between superconductivity and magnetism.
For example, the phase diagrams show an interplay between of an
itinerant spin density wave with superconductivity,
\cite{cruz,rotter1}
and density functional calculations show a proximity to
magnetism, with evidence for strong spin fluctuations.
\cite{singh-du,mazin1,yin,mazin2}
Other evidence for strong magnetic effects comes from
the unusual temperature dependence of the susceptibility, $\chi(T)$, which
for some compounds is increasing with $T$ up to high $T$,
\cite{klauss,mcguire}
and unusually large discrepancies between calculated and measured pnictogen
crystallographic positions. \cite{mazin2}
An important signature of the beyond mean field spin fluctuation effects
in the Fe-based superconductors is the fact that standard density functional
calculations give much more robust magnetic moments than experiment.

Comparing our experimental and calculated results for
BaCo$_2$As$_2$, we find a related situation. In particular, we do
not find clear evidence in transport or magnetization data for a
magnetic ordering (there is a change of behavior of the
magnetization at $\sim$ 60 K but the susceptibility continues
increasing below this temperature, and we do not find any
hysteresis). The implication is that the LDA yields a more
magnetic state than experiment. This generally occurs in materials
near a magnetic quantum critical point, where ordering is
suppressed by quantum fluctuations in the order parameter (i.e.
spin fluctuations). Examples of this behavior include the weak
itinerant ferromagnets ZrZn$_2$, Ni$_3$Al and the highly
renormalized paramagnets Ni$_3$Ga and Sr$_3$Ru$_2$O$_7$,
\cite{shimizu,kaul,mazin-z,aguayo,grigera} although the
renormalizations in those materials are apparently stronger than
in our measurements for BaCo$_2$As$_2$.

Further information comes from our single crystal measurements for
BaFeNiAs$_2$, which was prepared in a similar way to
BaCo$_2$As$_2$, using a near equal atomic mixture of Fe and Ni to
obtain approximately the same average $d$ electron count as
BaCo$_2$As$_2$. The single crystals of BaFeNiAs$_2$ were grown similar to that described above for BaCo$_2$As$_2$.
The measured specific heat $\gamma$
(Fig. \ref{NewFig3} is remarkably similar to those for
BaCo$_2$As$_2$. Therefore, at least for these properties, which
include the renormalization, the material behaves as a virtual
crystal, meaning that the alloy of Fe and Ni behaves in many respects as if it were a true pure compound of the element with the average atomic number (i.e. Co for an equal mixture). In this way, scattering is not strong enough to alter
the basic electronic structure, a fact that may be of importance
in understanding why superconductivity in the BaFe$_2$As$_2$ is
robust against the disorder introduced by alloying with Co or Ni.

To summarize we find from comparison of experimental data on single crystals
and LDA calculations that BaCo$_2$As$_2$ is a substantially renormalized
metal near a magnetic quantum critical point, probably of ferromagnetic
character. It will be of interest to tune the proximity to the critical point.
One interesting possibility would be to alloy on the
Ba-site with a magnetic element,
e.g. Eu, which may favor formation of a magnetically ordered state.

\acknowledgements

Work was supported by the Department of Energy, Division of
Materials Sciences and Engineering, Office of Basic Energy Sciences, U. S. Department of Energy.

\end{document}